\documentclass[twocolumn, switch]{article}
\usepackage{preprint}
\usepackage{hyperref}
\usepackage[numbers,square]{natbib}

\RequirePackage[labelformat=simple]{subcaption}                                
\hypersetup{colorlinks=true, linkcolor=purple, urlcolor=blue, citecolor=cyan, anchorcolor=black}

\usepackage[utf8]{inputenc}	
\usepackage[T1]{fontenc}
\usepackage{xcolor}
\usepackage{lineno}					
\usepackage{tikz} 					

\usepackage{placeins}

\usepackage{newfloat}
\DeclareFloatingEnvironment[name={Supplementary Figure}]{suppfigure}
\usepackage{sidecap}
\sidecaptionvpos{figure}{c}

\usepackage{titlesec}
\titlespacing\section{0pt}{12pt plus 3pt minus 3pt}{1pt plus 1pt minus 1pt}
\titlespacing\subsection{0pt}{10pt plus 3pt minus 3pt}{1pt plus 1pt minus 1pt}
\titlespacing\subsubsection{0pt}{8pt plus 3pt minus 3pt}{1pt plus 1pt minus 1pt}

\definecolor{lime}{HTML}{A6CE39}

\title{Predicting and forecasting reactivity and flux using long short-term memory models in pebble bed reactors during run-in} 

\shorttitle{Predicting and forecasting reactivity and flux using LSTM models in PBRs during run-in}

\usepackage{amssymb}
\usepackage{amsmath}

\usepackage{authblk}

\author
	[1,*]{Ian Kolaja}

\author[1]{Ludovic Jantzen}

\author[1]{Tatiana Siaraferas}

\author[1]{Massimiliano Fratoni}

\affil[1]{Nuclear Engineering Department, University of California Berkeley}
\affil[*]{Corresponding author email address: ikolaja@berkeley.edu}

\begin{document}

\twocolumn[\begin{@twocolumnfalse}

\maketitle


\begin{abstract}
Pebble bed reactor (PBR) operation presents unique advantages and challenges due to the ability to continuously change the fuel mixture and excess reactivity. Each operation parameter affects reactivity on a different timescale---for example, fuel insertion changes may take months to fully propagate, whereas control rod movements have immediate effects. In-core measurements are further limited by the high temperatures, intense neutron flux, and dynamic motion of the fuel bed. In this study, long short-term memory (LSTM) networks are trained to predict reactivity, flux profiles, and power profiles as functions of operating history and synthetic batch-level pebble measurements, such as discharge burnup distributions. The model’s performance is evaluated using unseen temporal data, achieving an $R^2$ of 0.9914 on the testing set. The capability of the network to forecast reactivity responses to future operational changes is also examined, and its application for optimizing reactor running-in procedures is explored.
\end{abstract}


\vspace{0.5cm}

\end{@twocolumnfalse}]

\section{Introduction}

The operation of pebble bed reactors (PBRs) poses additional advantages and challenges not realized by reactors with static fuel assemblies. Because PBR operators have control over the fuel that is inserted into the core, they can control its excess reactivity. However, fuel pebbles inserted into the reactor core can take months to fully move through it. For example, in the generic fluoride salt-cooled high temperature reactor (gFHR) benchmark model published by Kairos Power~\cite{gfhr}, the pebbles usually take 60-70 days to pass through the core fully. This means the impact on reactivity has significant latency with respect to fuel insertion changes. This is particularly relevant for reactor running-in phase, which generally involves dummy graphite pebbles and/or other enrichment fuel~\cite{run-in-phase-fhr-kairos-kang}. The ability to understand the long-term consequence of changing reactor parameters is crucial for safely operating the reactor. This is especially true for running-in phase, where operators want to minimize the amount of time spent a low power while keeping pebbles within safety and fuel qualification limits.

Furthermore, direct core measurement is much more limited for PBRs compared to light water reactors (LWRs) due to their higher operating temperatures. Presently, no commercially available neutron flux sensors can reliably function above 550°C~\cite{advanced-reactors-sensor-summary-korsah}. Direct in-core temperature measurements are also limited, with only the surrounding graphite being measurable using thermocouples embedded in the reflector blocks. Thermocouples are also prone to drift in high temperature high flux environments~\cite{advanced-reactors-sensor-summary-korsah}. Thus, most of the information available to the operator lies in the operation history and the properties of the discharged fuel. It is possible to accurately predict the burnup, fuel composition, fast and thermal fluence, and average radial path of a pebble using its measured gamma spectrum~\cite{kolaja2025burnupmeasurementusingbent}. By aggregating discharge pebble data and tracking its evolution over time, information about the core can be determined. 

Accurately predicting how reactivity is affected by these different time-dependent effects lends itself to time series analysis methods and machine learning models. Recurrent neural networks (RNNs) are deep learning models that can process sequential input data~\cite{rnn-review-lipton}. While commonly associated with language processing, they are very suitable for time series regression. One class of RNN is a long short-term memory (LSTM) network, which feature a memory cell with input, forget, and output gates that control what information the model has access to~\cite{deep-learning-goodfellow}. The ability to train the model to selectively retain certain information allows it to make predictions based on the history of reactor operation and measurement. This enables the prediction of reactivity and the flux and power profiles as they evolve over time. This work discusses the generation of core operation sequence data using a zone-based depletion model, the implementation of an LSTM model, and an assessment of its ability to predict the short and long term behavior of reactivity and the core flux distribution in a variety of situations. This model is then used to steer the zone simulator through running-in phase, iterating between data generation and re-training to optimize the process. This work is based on Chapter~3 of the author’s PhD dissertation~\cite{ian-dissertation}, completed under the supervision of Massimiliano Fratoni. Key data and analysis code used in this work have been archived on Zenodo~\cite{zenodo_pearlsim}.

\begin{figure}[htbp]
\centering
  \begin{subfigure}[b]{0.75\columnwidth}
    \centering
  \includegraphics[width=\textwidth]{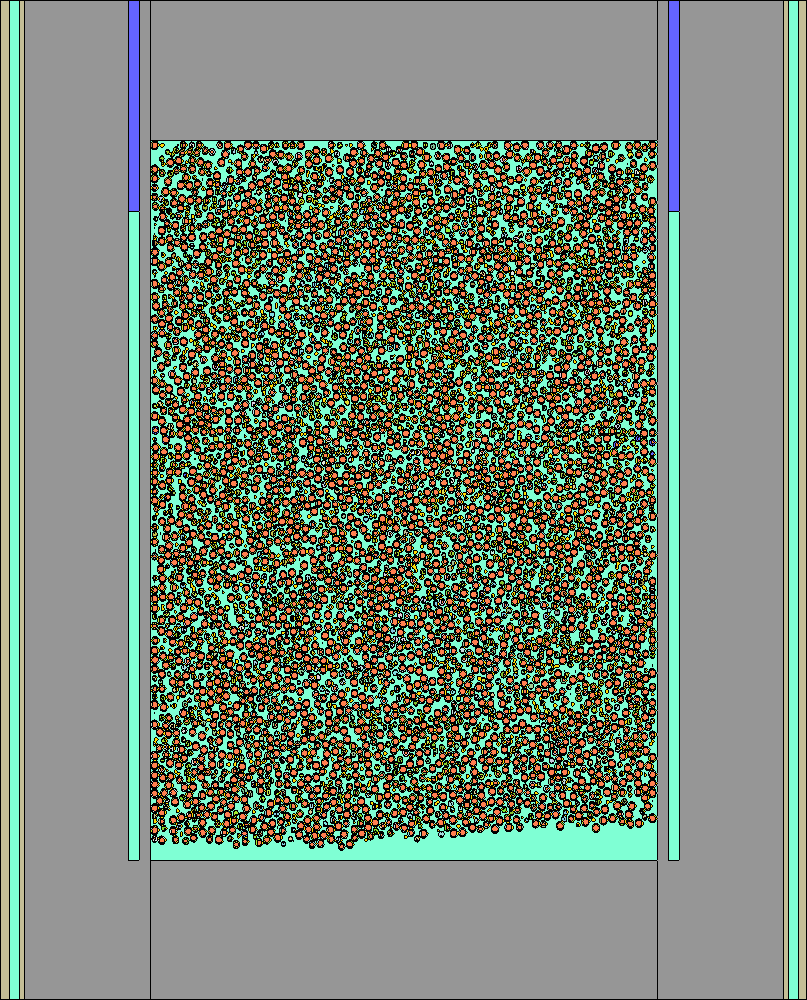}
  \label{fig:gfhr_model:parta}
  \end{subfigure}
  \begin{subfigure}[b]{0.4\columnwidth}
    \centering
    \includegraphics[width=\textwidth]{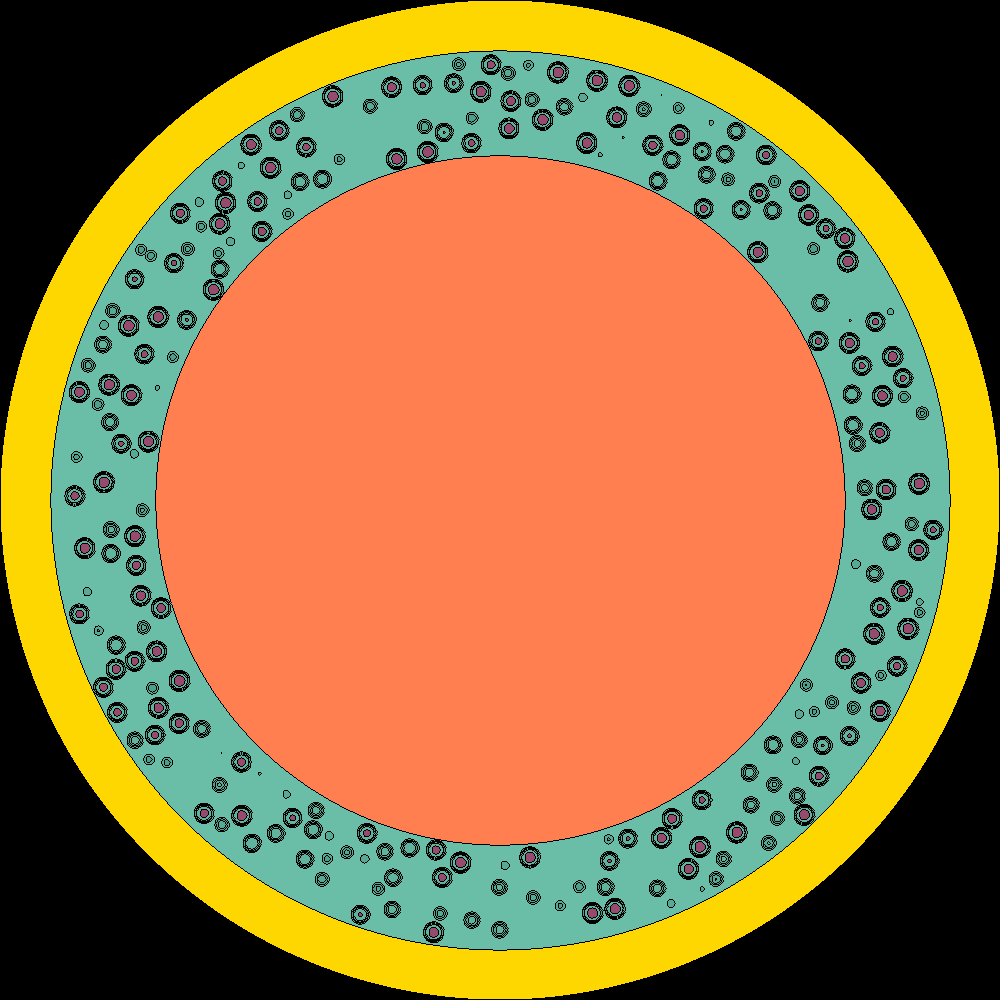}
  \label{fig:gfhr_model:partb}
  \end{subfigure}
  \caption{Serpent-generated plot of the gFHR model implemented in Serpent, showing the full core (top) and fuel pebble (bottom).}
  \label{fig:gfhr_model}
\end{figure}

\section{Data Generation}
\label{sec:DataGeneration}

\subsection{Zone Based Core Simulation with PEARLSim}
\label{subsec:PEARLSim}
The pebble-explicit advanced reactor learning simulator (PEARLSim) tool is a Serpent 2.2.0~\cite{serpent} wrapper that simulates a PBR core with a combination of Monte Carlo particle transport, depletion methods, and high level fuel inventory management~\cite{zenodo_pearlsim}. Similar to the depletion method demonstrated in the gFHR benchmark~\cite{gfhr}, the active core volume is divided into radial and axial zones. Rather than uniquely tracking every pebble, large groups of pebbles are collectively identified based on the geometric zone they occupy as well as their relative burnup. For this study, the gFHR design is modeled and divided into 4 radial zones, 10 axial zones, and up to 12 burnup groups. This means up to 480 unique fuel pebble materials are tracked in Serpent. The pebble locations are explicitly defined based on the result of a discrete element motion simulation. The TRISO locations within a pebble matrix are determined using Serpent's disperse routine. A plot of the gFHR geometry is shown in Figure~\ref{fig:gfhr_model}. A handful of key fuel parameters from the benchmark are listed in Table~\ref{tab:gFHR_data}.

\begin{table}[htbp]
    \centering
    \small
    \begin{tabular}{|cc|}
    \hline
    \multicolumn{2}{|c|}{\textbf{Core (cylinder)}}                                                                                                                              \\ \hline
    \multicolumn{1}{|l|}{\textbf{\begin{tabular}[c]{@{}c@{}}Total power\end{tabular}}}                                                   & 280 MW                                                                       \\ \hline
    \multicolumn{1}{|c|}{\textbf{\begin{tabular}[c]{@{}c@{}}Number of \\ pebbles\end{tabular}}}  & 250 190                                                                      \\ \hline
    \multicolumn{2}{|c|}{\textbf{Pebbles}}                                                                                                                                      \\ \hline                                               
    \multicolumn{1}{|c|}{\textbf{\begin{tabular}[c]{@{}c@{}}Layers radius\\ (cm)\end{tabular}}}  & 1.38, 1.8, 2                                                                   \\ \hline
    \multicolumn{2}{|c|}{\textbf{TRISO Particles}}                                                                                                                              \\ \hline
    \multicolumn{1}{|c|}{\textbf{\begin{tabular}[c]{@{}c@{}}Lattice \\ structure\end{tabular}}}  & dispersed (PF = 22\%)                                                        \\ \hline
    \multicolumn{1}{|c|}{\textbf{\begin{tabular}[c]{@{}c@{}}Layers radius\\ ($\boldsymbol{\mu}$m)\end{tabular}}} & \begin{tabular}[c]{@{}c@{}}212.5, 312.5, 352.5, \\ 387.5, 427.5\end{tabular} \\ \hline
    \multicolumn{1}{|c|}{\textbf{Fuel form}}                                                          & UCO                                                                          \\ \hline
    \multicolumn{1}{|c|}{\textbf{Enrichment}}                                                    & 19.55 wt\%                                                                   \\ \hline
    \end{tabular}
    \caption{Key design and operating parameters from the Kairos Power gFHR benchmark~\cite{gfhr}.}
    \label{tab:gFHR_data}
\end{table}

A Serpent input is generated for each time step. Pebble shuffling occurs by randomly assigning fuel materials to pebble locations in each zone in ratios proportional to the zone's fuel inventory. One iteration captures the depletion that happens over the user defined timestep, which can be varied between steps and is directly related to the overall circulation rate in the core. In the gFHR benchmark, the 522 day residence time with 10 axial zones and 8 passes corresponds to a depletion time step of 6.525 days and a circulation rate of 3,800 pebbles/day.

\begin{figure}[htbp]
  \centering
  \includegraphics[width=0.9\columnwidth]{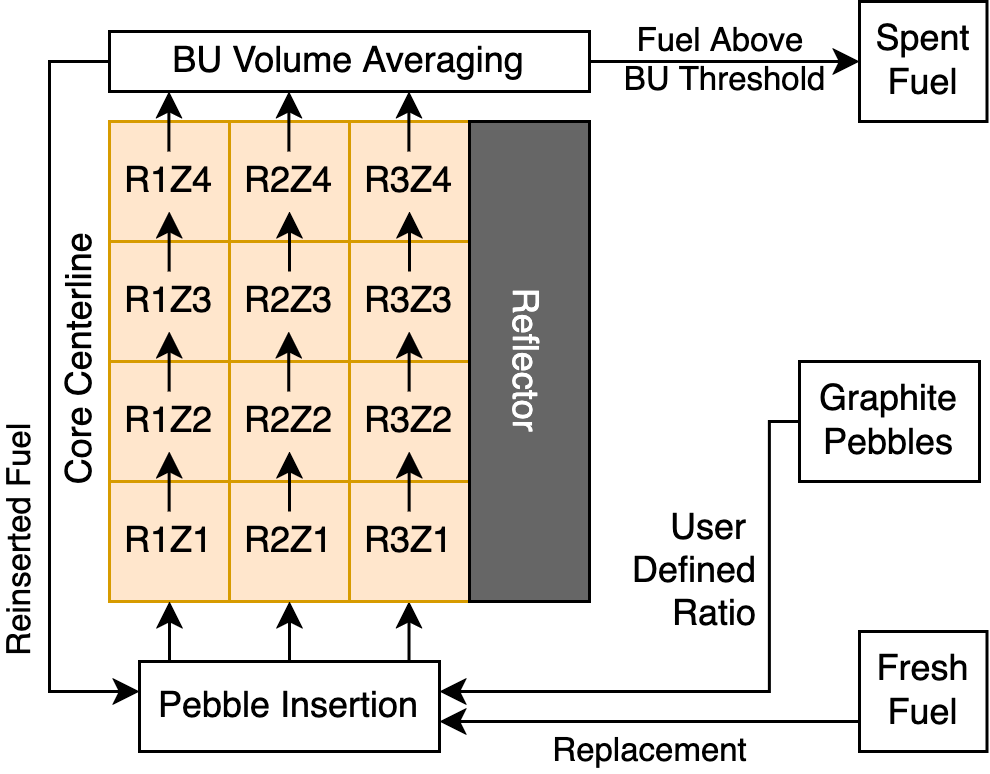}
  \caption{Schematic of core simulator zones, as well as the flow of pebbles through them. This schematic only has 3 radial zones and 4 axial zones.}
  \label{fig:zone_diagram}
\end{figure}

On every iteration, the upward motion of pebbles due to buoyancy is simulated by moving the pebble inventory information from one zone to the zone above. Pebbles in the top zones are discharged from the core, and they are reassigned to new burnup groups for volume averaging. The movement of pebbles is shown in Figure~\ref{fig:zone_diagram}. The burnup groups are defined with evenly spaced bins. The spacing of these bins is adjusted until there are 12 non-empty burnup bins occupied by discharged fuel materials. Then, for each burnup group $g$, a weighted average of the nuclide concentrations of its constituent pebbles is taken according to Equation~\ref{eqn:volumeaverage}. 
\begin{equation}
  \bar{C}^x_g = \sum \frac{N_i}{N_g}C^x_i
  \label{eqn:volumeaverage}
\end{equation}
For each pebble type in the group, the concentration of nuclide $x$, $C_x^i$, is weighted by the number of pebbles of that type $N_i$ over the total number of pebbles in the group $N_g$.  

Discharged fuel groups are either reinserted or discarded based on their burnup. Materials with burnup $\bar{b_g}$ exceeding the user specified burnup threshold, $T$, are partially or fully removed from the core. The burnup $B$ of pebbles in the group is assumed to be normally distributed, $B \sim N(\mu, \sigma^2)$, where $\mu=\bar{b_{g}}$ and $\sigma=c\bar{b_{g}}$. Thus, the fraction of pebbles, $f_{removed}$ removed is given by Equation~\ref{eqn:threshold_fraction}.
\begin{equation}
  f_{removed} \equiv P(B<T)
  \label{eqn:threshold_fraction}
\end{equation}
The remaining volume averaged, burnup-binned fuel groups are then reinserted at the bottom of the core on the next iteration. Removed pebbles are replaced by fresh fuel.

The value of $c$ was set to make $c\bar{b_g}$ match the distribution of discarded pebbles simulated with HxF. HxF is another tool that uses Serpent 2.2.0 that is capable of simulating the full irradiation history of individual pebbles without zone assumptions~\cite{HxF}. While HxF generates the highest fidelity of pebble data available, it is far too computationally costly to generate the amount of operation data needed for time series models, and is only used to inform the distribution of pebble burnup. The gFHR model was simulated to equilibrium at 100\% power with a 180 MWd/kgHM discard threshold. The resulting burnup distribution of discarded pebbles was used to determine a value of $c=0.02288$.

There are some departures from the benchmark model. The radial zone boundaries are defined to impose equal zone volume rather than to increase linearly. While the original gFHR benchmark grouped pebbles by their number of passes through the core, PEARLSim uses burnup. The insertion of "dummy" graphite pebbles is also explored in this study. Finally, simple control rods are implemented, being inserted in the side reflector from the top of core. The rods are made of boron carbide with 100\% $^{10}$B and can be inserted all the way to the bottom of the active core volume. This allows the study to explore a more extensive list of operating parameters than the power and circulation rate, which are also varied from their benchmark values of 280 MW and 3,800 pebbles/day. 

PEARLSim can be used to model cores with conic regions as well. The radial boundaries can be defined with piecewise functions rather than constant values. Each radial channel can also feature a different number of axial zones, changing the bulk vertical velocity of pebbles in those axial zones. While not used in this study, this functionality is important to note for using the simulator to generate data for more detailed designs that would feature a wider range of pebble flow velocity, particularly near the reflector.

\subsection{PEARLSim Output and ML Feature Design}

\begin{figure*}[ht]
\centering
\begin{subfigure}[b]{0.3\textwidth}
    \centering
    \includegraphics[width=\textwidth]{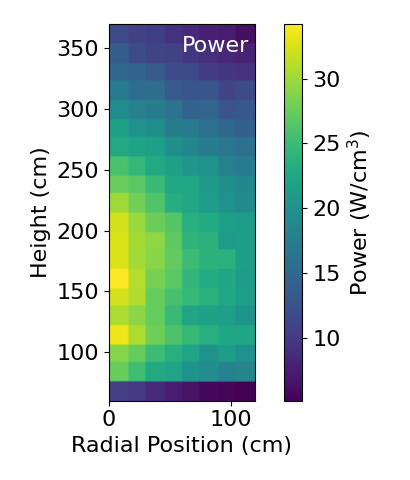}
\end{subfigure}
\begin{subfigure}[b]{0.3\textwidth}
    \centering
    \includegraphics[width=\textwidth]{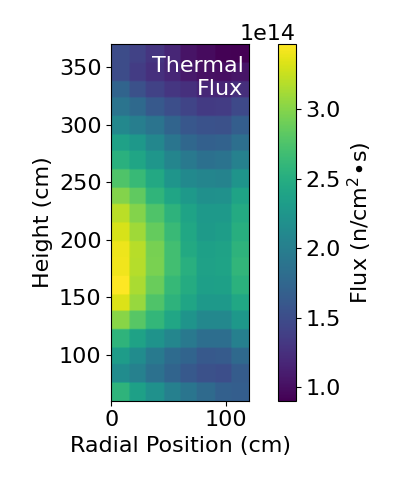}
\end{subfigure}

\begin{subfigure}[b]{0.3\textwidth}
    \centering
    \includegraphics[width=\textwidth]{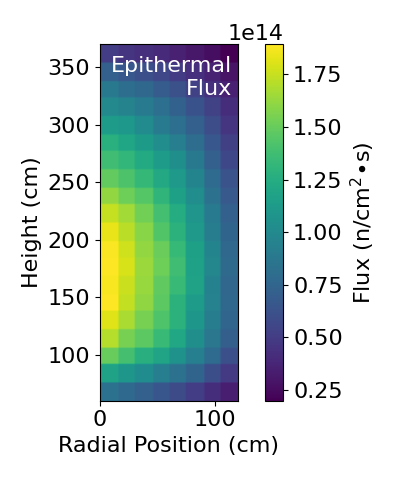}
\end{subfigure}
\begin{subfigure}[b]{0.3\textwidth}
    \centering
    \includegraphics[width=\textwidth]{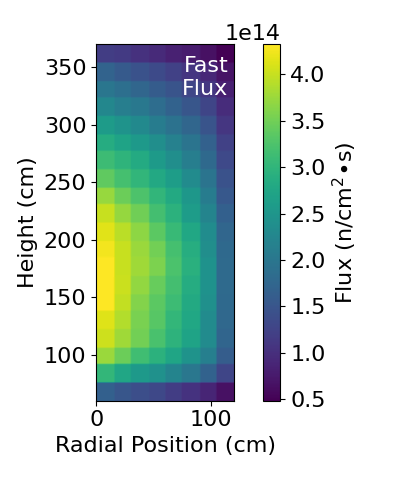}
\end{subfigure}
\caption{Examples of the 20 by 8 subzone meshes generated by PEARLSim for power (top left), thermal flux (top right), epithermal flux (bottom left), and fast flux (bottom right). The power mesh typically has a relative uncertainty of 1.8\%, while the flux mesh has 1.3\%.}
\label{fig:mesh_examples}
\end{figure*}

For a given operation sequence, PEARLSim calculates the group-averaged pebble composition data, $k_{eff}$, and the power and flux meshes. Its results can be expanded to include any parameter calculated by Serpent. The tally meshes have 8 radial subdivisions and 20 axial subdivisions that are linearly spaced. There are three energy groups for thermal, epithermal, and fast neutrons, which are separated at 0.625 eV and 1 keV. Examples of volume averaged meshes at equilibrium are shown in Figure~\ref{fig:mesh_examples}. Reactivity was calculated from $k_{eff}$ using Equation~\ref{eqn:reactivity} and used as a target variable. 
\begin{equation}
  \rho=\frac{k_{eff}-1}{k_{eff}}
  \label{eqn:reactivity}
\end{equation}
The meshes were subject to dimensionality reduction, as is explained in Section~\ref{subsec:flux_power_mesh_dimension_reduction}. To balance performance and accuracy, 30 cycles (10 inactive) with 24,000 histories were simulated on every depletion step.

The operation sequence include user defined values for the fraction of dummy graphite pebbles inserted, the total power for normalization, the control rod insertion, the circulation rate (implemented through varying the depletion time step, which is used interchangeably in this work), and the burnup threshold for discarding pebbles. It is assumed that an operator has perfect knowledge of these variables. The average power per pebble is also calculated by dividing the current power of the reactor by the number of fuel pebbles in the reactor. The number of fuel pebbles can be determined indirectly by knowing fuel insertion history.

Next, observable features were computed from the discharge pebbles. These features serve as dependent input variables to the model. While measuring every pebble as it comes out of the core is feasible, directly inputting pebble-wise data into an LSTM would likely fail to capture coherent trends over time while massively increasing the size of the network. Instead, a window of time can be used in which features are computed from all pebbles discharged during that period. For simplicity, this measurement period was assumed to match the depletion time step, which corresponds to batch measurements of about 25,000 pebbles. For this window to be made smaller, PEARLSim would need to be run with more axial zones, or a pebble-wise depletion tool like HxF would need to be run. Both options increase the computation time without a clear benefit in representing long-term trends in average core behavior. 

\begin{figure}[htbp]
  \centering
  \includegraphics[width=0.9\columnwidth]{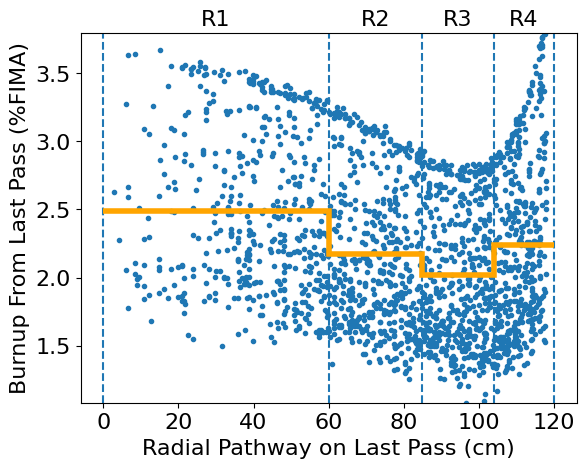}
  \caption{Demonstration of how last-pass burnup would be averaged for radial zones using HxF equilibrium data as an example. This shows how pebble-wise measurements can be spatially binned to inform the model.}
  \label{fig:burnup_radial_averaging}
\end{figure}

Previous work has shown the possibility of measuring discharge fuel gamma spectra to predict the average radial pathway of a pebble on its most recent pass through the core, as well as its total and last-pass burnup~\cite{kolaja2025burnupmeasurementusingbent}. By grouping pebbles according to their predicted path, their predicted values for burnup and last-pass burnup can be spatially binned. This is illustrated in Figure~\ref{fig:burnup_radial_averaging} by using data from an HxF simulation to show how real measurements could be zone averaged. It is assumed that using PEARLSim zone averages for burnup are roughly equivalent to the results of this binning process, removing the need to run HxF to get representative values. 

Four features related to burnup have been created. First, the average last-pass burnup of pebbles, $B^{last}_r$, discharged from each radial zone was calculated using Equation~\ref{eqn:radialburnup} (in units of FIMA).
\begin{equation}
  B^{last}_r=\frac{\sum^G_{g=1}N_{r,g}B^{last}_{r,g}}{N_r}
  \label{eqn:radialburnup}
\end{equation}
This was done using the number of pebbles $N_{r,g}$ and their average last pass burnup $B^{last}_{r,g}$ for each burnup group $g$ in the discharge radial zone $r$. These values correlate with the amount of flux in the corresponding radial zones, making these features potentially useful for reconstructing the flux and power meshes.

\begin{figure}[ht]
\centering
\begin{subfigure}[b]{0.9\columnwidth}
    \centering
    \includegraphics[width=\textwidth]{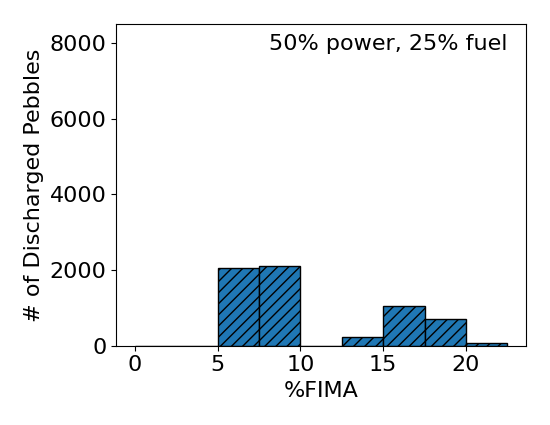}
    \label{fig:burnup_binning:parta}
\end{subfigure}
\begin{subfigure}[b]{0.9\columnwidth}
    \centering
    \includegraphics[width=\textwidth]{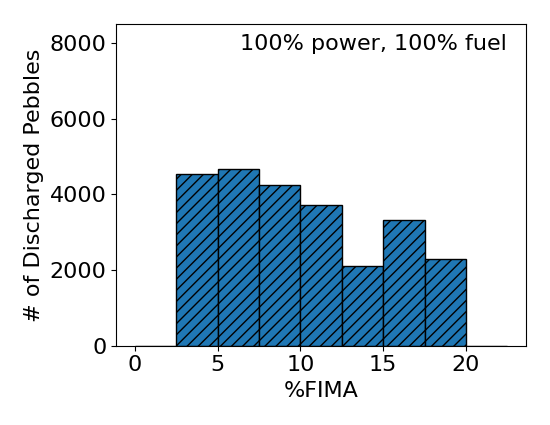}
    \label{fig:burnup_binning:partb}
\end{subfigure}
\caption{Example of discharge pebble burnup binning for a core at low power early in operation (top) and at full power close to equilibrium (bottom).}
\label{fig:burnup_binning}
\end{figure}

The numbers of pebbles discharged during a timestep that fall into 9 total burnup bins between 0 \%FIMA and 22.5 \%FIMA are counted, $N_b$. This coarsely captures the distribution of burnup across all pebbles. This is shown for two different core states in PEARLSim in Figure~\ref{fig:burnup_binning}. The average total burnup of pebbles discharged during a timestep is also directly calculated, $\bar{B}^{total}$ which should have a clear relationship with reactivity. Finally, the number of pebbles that are discarded due to exceeding the burnup threshold, $D$, is tracked.

Some variance is applied to the dependent input variables to emulate potential sources of error. This includes counting uncertainty associated with short gamma measurement times and the lower reported accuracy of pebble pathway predictions~\cite{kolaja2025burnupmeasurementusingbent}. It also makes the model more robust and prevents overfitting on patterns of overly precise values. The variance applied to each feature is shown in Table~\ref{tab:noise_table}, and corresponds to a Gaussian with standard deviation calculated with Equation~\ref{eqn:guassian_noise}.
\begin{equation}
  \sigma=\frac{MAPE}{100\sqrt{2/\pi}}
  \label{eqn:guassian_noise}
\end{equation}

\begin{table}[!htb]
    \centering
    \begin{tabular}{|c|c|}
    \hline
    \textbf{Parameter}                        & \textbf{MAPE applied} \\ \hline
    \textbf{Burnup Bins}                      & 5\%                   \\ \hline
    \textbf{Average Discharge Burnup}         & 2.5\%                 \\ \hline
    \textbf{Discarded Pebbles}                & 5\%                   \\ \hline
    \textbf{Avg. Last Pass BU by Radial Zone} & 10\%                  \\ \hline
    \end{tabular}
    \caption{Mean Absolute Percent Error (MAPE) applied to each dependent input variable.}
    \label{tab:noise_table}
\end{table}

\subsection{Flux and Power Mesh Dimension Reduction}
\label{subsec:flux_power_mesh_dimension_reduction}

As shown in Figure~\ref{fig:mesh_examples}, the meshes for flux and power have high dimensionality, but their shape changes very little over the course of operation. Modeling each cell individually would be computationally expensive and largely redundant. Dimensionality reduction techniques could instead be used to reduce the number of variables while maintaining as much information as possible. Principal component analysis (PCA) was selected to linearly transform the mesh data into a much smaller set of independent variables. Although more sophisticated approaches, such as autoencoders, can capture nonlinear relationships in the data and often outperform PCA~\cite{reducing-dimensionality-nn-autoencoder-hinton}, they were not used here for several reasons. Autoencoders require a considerably larger amount of data~\cite{pca-autoencoders-low-data-al-digeil} to perform well. They also suffer from worse run-time by up to two orders of magnitude~\cite{comparison-between-autoencoders-dim-reduction-fournier}. Finally, the need to tune the hyperparameters for an additional neural network and ensure stable convergence would introduce significant overhead that is unnecessary to demonstrate the capabilities of the LSTM. For these reasons, PCA was chosen as a simple, fast, and physically interpretable method for reducing mesh dimensionality for this proof of concept work.

\begin{figure}[htbp]
\centering
\begin{subfigure}[b]{0.49\columnwidth}
    \centering
    \includegraphics[width=\textwidth]{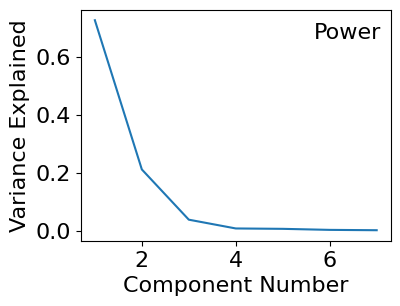}
    \label{fig:scree_plots:parta}
\end{subfigure}
\begin{subfigure}[b]{0.49\columnwidth}
    \centering
    \includegraphics[width=\textwidth]{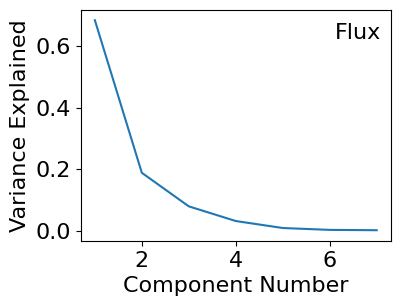}
    \label{fig:scree_plots:partb}
\end{subfigure}
\caption{Scree plots showing the amount of variance captured by increasing the numbers of principal components for the power meshes (left) and flux (right).}
\label{fig:scree_plots}
\end{figure}

It is important to only model statistically significant principal components. Extraneous higher order components are likely to only capture noise. Explained variance ratios, which are plotted in Figure~\ref{fig:scree_plots}, can be used to select a useful number of components. These ratios quantify how much of the variance in the original distribution is captured by each component. The typical flux and power meshes produced in this study have an average relative Monte Carlo uncertainty of 1.8\% and 1.3\% respectively. Thus, it is reasonable to start with a number of components such that the cumulative explained variance ratio for each mesh reaches 98.2\% and 98.7\% respectively. This is achievable with 5 components for both meshes. It can be verified whether the higher order components are useful by checking if the accuracy is improved by setting them to zero during reconstruction.

\begin{figure*}[ht]
\centering
\begin{subfigure}[b]{0.43\textwidth}
    \centering
    \includegraphics[width=\textwidth]{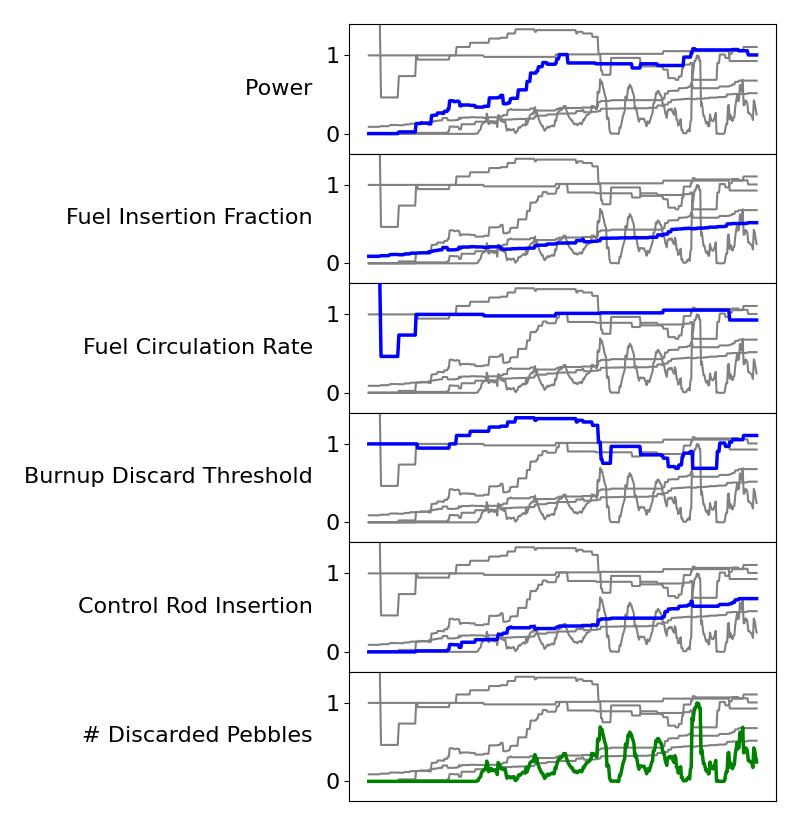}
    \label{fig:random_sequence_example:parta}
\end{subfigure}
\begin{subfigure}[b]{0.43\textwidth}
    \centering
    \includegraphics[width=\textwidth]{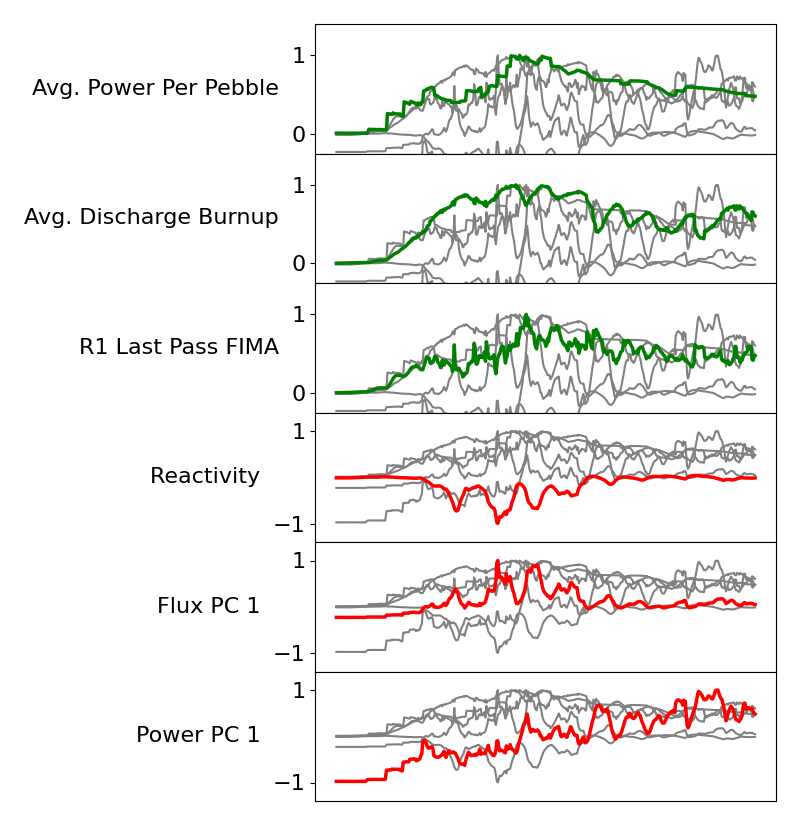}
    \label{fig:random_sequence_example:partb}
\end{subfigure}
\caption{Example of a randomly generated operation sequence in PEARLSim, showing the input variables (blue), dependent input variables (green) and target variables (red). All values are normalized to their nominal equilibrium values.}
\label{fig:random_sequence_example}
\end{figure*}

\subsection{Sequence Generation with PEARLSim}
\label{subsec:SequenceGeneration}
PEARLSim was used to simulate a wide range of operational sequences including both hand-crafted and randomly generated simulations. Some of the hand-crafted sequences emulate a reasonable power ascension sequence, starting from low power and a high fraction of graphite pebbles and progressively moving toward equilibrium. Other hand-crafted sequences focus entirely on one or two operating parameters, providing data that shows how different combinations of controls are related. 14 handcrafted sequences and 19 randomized sequences were generated for use in the model.

When generating random sequences, each input parameter has step sizes and probabilities assigned for whether it is varied and whether it is changed positively or negatively. The magnitude of each parameter perturbation and the length that the generated control vector is maintained is also randomized. Each random sequence has different probabilities to capture a broader, less human-biased subset of the overall operating domain. An example of a randomly generated is shown in Figure~\ref{fig:random_sequence_example}.

LSTMs require fixed-length subsets of the data. Accordingly, each simulated time series dataset was reformatted into a three-dimensional array with dimensions corresponding to the number of samples, the number of time steps per sample, and the number of features. Each sample therefore represents the temporal evolution of all features over a specified time window. The length of this window (i.e., the number of depletion time steps) must be chosen to balance computational efficiency with the model’s capacity to learn long-term temporal relationships. Because there are 10 axial zones, a window size around 10 steps is expected to perform the best, since that is the minimum amount of time it takes for fuel insertion changes to fully propagate through the core. Through manual tuning, a window size of eight was found to perform well. This is sufficiently long to capture delayed operational effects while avoiding excessive sequence memorization and overfitting.

\section{Model Training and Assessment}

\subsection{LSTM Training and Forecasting}
\label{subsec:LSTMtraining}

\begin{figure}[hb]
  \centering
  \includegraphics[width=0.99\columnwidth]{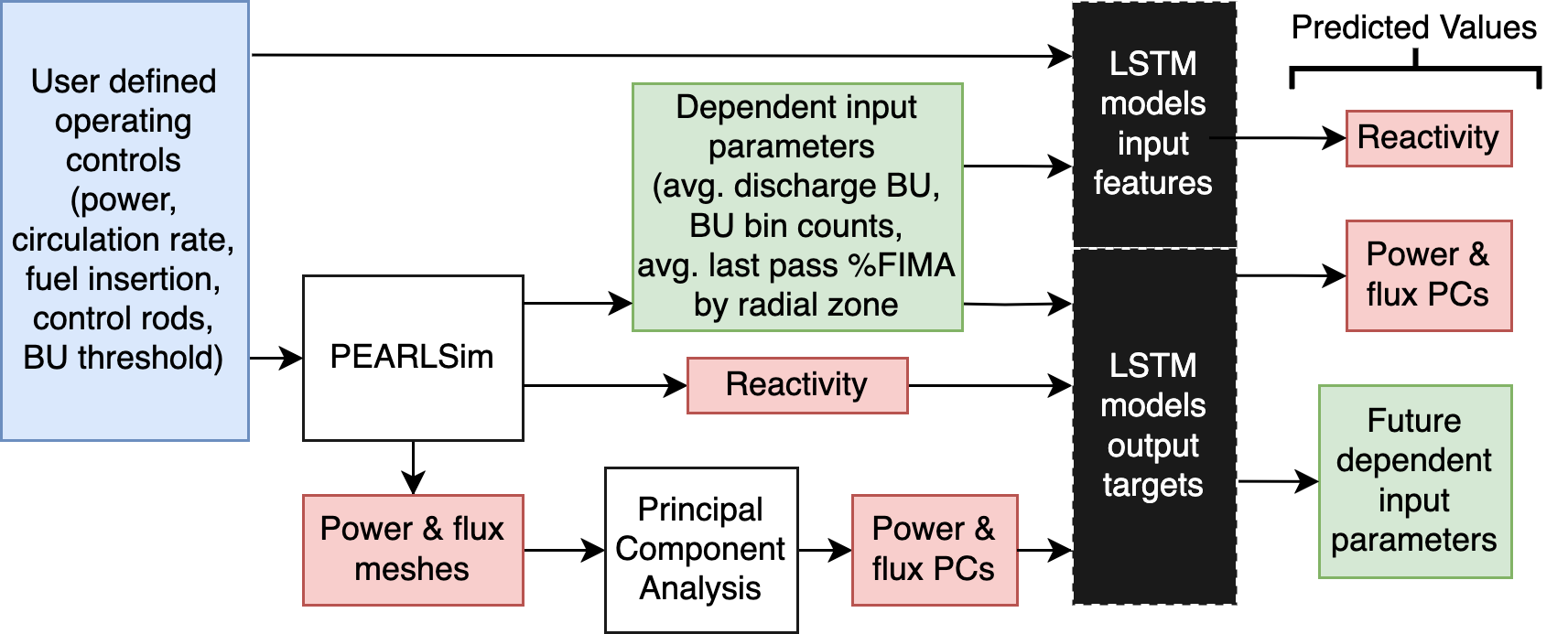}
  \caption{Conceptual overview of LSTM inputs and outputs. Each target variable has its own model trained for it.}
  \label{fig:lstm_diagram}
\end{figure}

The LSTM implementation by Keras~\cite{keras} was used for this work. One handcrafted sequence was entirely held out, while the rest were used in training. 70\% of the samples were used in training, 15\% were used for validation, and 15\% were used for testing. A separate network was trained for each output parameter, including excess reactivity and the five principal components of both flux and power. In addition, the model was trained to predict the values of dependent input variables, such as average discharge burnup, at the subsequent time step. The general network architecture is shown in Figure~\ref{fig:lstm_diagram}.

\begin{figure}[ht]
\centering
\begin{subfigure}[b]{0.82\columnwidth}
    \centering
    \includegraphics[width=\textwidth]{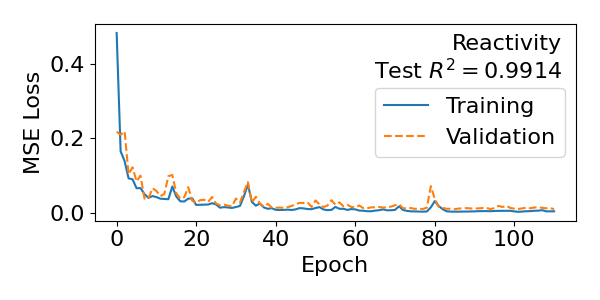}
\end{subfigure}
\begin{subfigure}[b]{0.82\columnwidth}
    \centering
    \includegraphics[width=\textwidth]{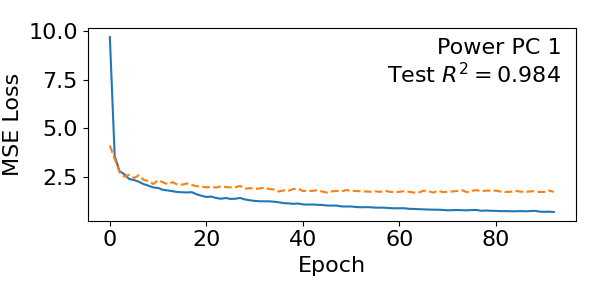}
\end{subfigure}
\begin{subfigure}[b]{0.82\columnwidth}
    \centering
    \includegraphics[width=\textwidth]{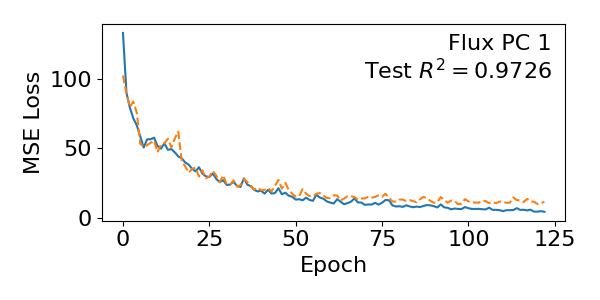}
\end{subfigure}
\begin{subfigure}[b]{0.82\columnwidth}
    \centering
    \includegraphics[width=\textwidth]{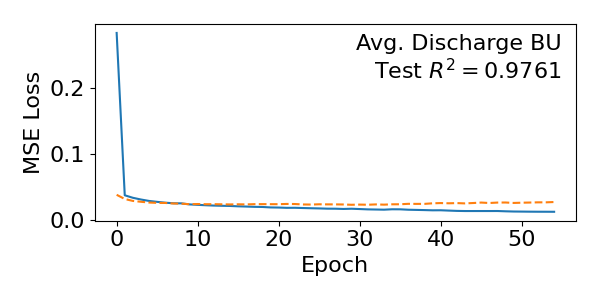}
\end{subfigure}
\caption{Learning curves for a handful of parameters. The curves extend far beyond when an optimal set of weights is found due to the large early stopping patience.}
\label{fig:learning_curves}
\end{figure}

By predicting the dependent input features on the next time step, it becomes possible to predict the core state an arbitrary number of time steps into the future. An operator could also modify future operating parameters. This capability allows operators to ensure that the reactor will remain critical even after changes in pebble insertion have propagated fully through the core. However, this requires a lot of training data to be fully reliable, as the model gets progressively less accurate the further into the future it is used.

The Adaptive Moment Estimation (Adam) optimizer was used in conjunction with a learning rate schedule~\cite{adam-optimizer-introduction-kingma}. The initial learning rate was set to 0.01 and decayed exponentially by a factor of 0.9 every 10,000 steps~\cite{deep-learning-goodfellow}. Early stopping based on validation loss was used to determine the optimal number of training epochs. Training was run for a minimum of 10 epochs and terminated when the validation loss did not improve for 25 consecutive epochs. The model weights corresponding to the lowest validation loss were retained. 

Although the input window size could, in principle, be optimized for each target, doing so would substantially complicate the workflow. Therefore, a fixed window size of eight time steps was used, as it performed well for reactivity predictions. L2 regularization was applied to penalize large weights, and recurrent dropout with a probability of 10\% was employed to improve generalization and ensure that the model emphasized physical trends~\cite{rnn-regularization-zaremba, deep-learning-goodfellow}. The number of layers as well as their unit sizes were optimized for each target variable using k-fold cross validation with 5 folds. The optimal network sizes for each target variable are shown in Table~\ref{tab:Rsquared_values_and_hyperparams}. Representative learning curves for selected output variables using the full training set after hyperparameter tuning was performed are shown in Figure~\ref{fig:learning_curves}.

\subsection{Feature Importance}
\label{subsec:feature_importance}

Assessing the usefulness of different input features for accurate predictions can help eliminate redundant features, understand how the model makes predictions, and emphasize what measurements are important to inform the model. Quantifying feature importance is not as straightforward in LSTMs as it is in other types of ML models, such as those that use decision trees. Retraining the model with every combination of features has a very high computational cost. Instead, the permutation importance is quantified by randomly shuffling the values of an input parameter across the full dataset. The Absolute Mean Error (MAE) is compared before and after the input parameter is shuffled for each target variable, showing the loss of accuracy when the model loses any useful information about the input.

\begin{figure*}[p]
\centering
\includegraphics[width=0.98\textwidth]{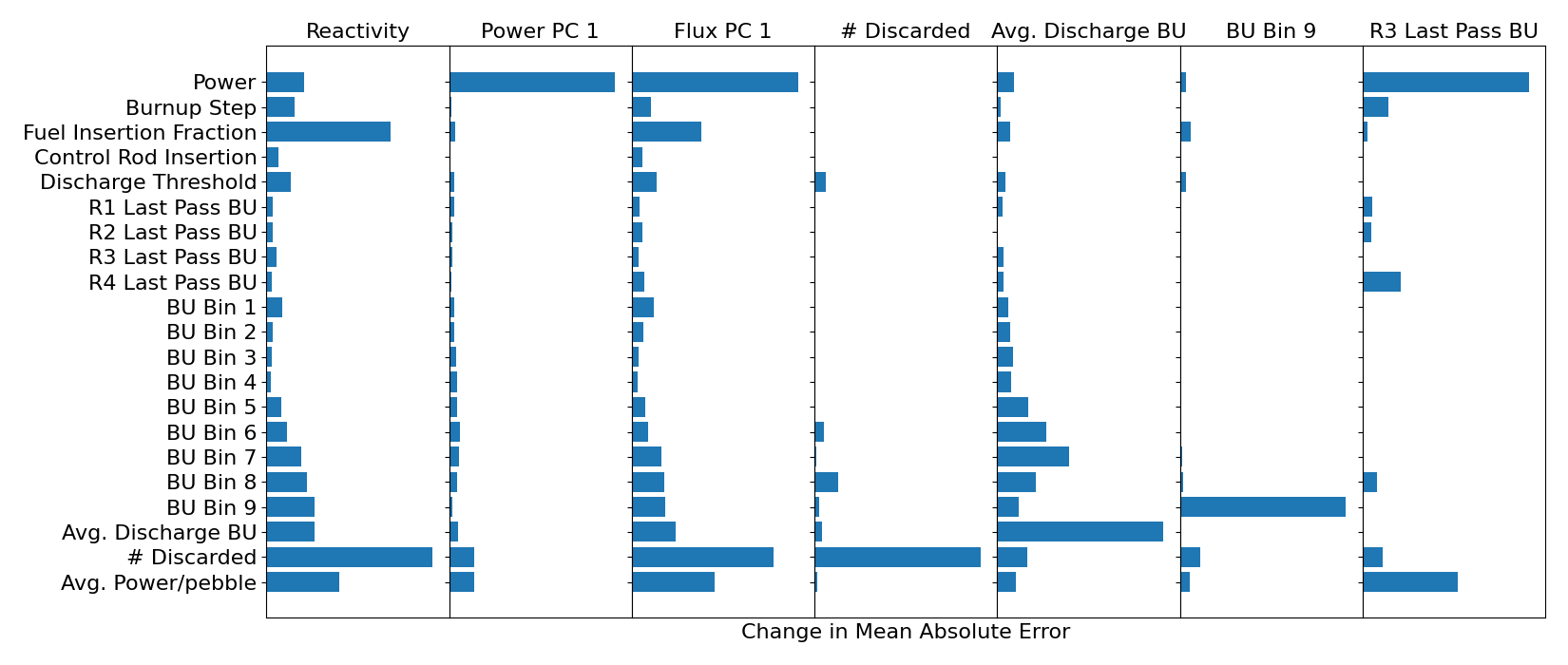}
\caption{LSTM permutation feature importance for key target variables and future values of dependent input features.}
\label{fig:feature_importance_targets}
\end{figure*}

The permutation importance of each feature is shown in Figure~\ref{fig:feature_importance_targets} for key output variables. For reactivity, the most important features included fuel insertion, average discharge burnup, and the number of discarded pebbles. The high importance of the number of discarded pebbles and burnup bins 7-9 suggests that the model is capable of identifying when many pebbles are about to reach the discard threshold. This is often followed by a spike in reactivity as an influx of replacement fresh fuel is consequently inserted at the bottom of the core.

Some sets of input variables are collinear, which reduces the loss of accuracy observed when just one of them is shuffled. The model still has access to most of the same information in these cases. This is especially apparent with the last-pass burnup variables, although the relatively low importance of them as a group could motivate condensing them into one variable. The radial deviation appears to provide less information than was expected. Furthermore, the path that the pebbles took through the core is relatively difficult to accurately predict, making the grouping of the pebbles into radial zones less reliable~\cite{kolaja2025burnupmeasurementusingbent}. 

The future dependent input features shown a unique set of trends. Unsurprisingly, they are largely sensitive to their current value, except for the last-pass radial burnup values. These are instead sensitive to the overall power and power per pebble in the core, which makes sense since they are much more related to short-term conditions rather than long-term pebble history.

Although not shown, the higher rank principal components are increasingly more sensitive to the number of discarded pebbles as they capture smaller perturbations in the flux and power distributions due to oscillations in the fuel inventory. 

\begin{table*}[p]
\caption{Coefficient of determination for target variables when using the model on the training and testing datasets respectively.}
\begin{tabular}{|c|c|c|l|ccccl}
\cline{1-4} \cline{6-9}
\textbf{Parameter}         & \textbf{Train $R^2$} & \textbf{Test $R^2$} & \textbf{\begin{tabular}[c]{@{}l@{}}Best Hidden\\ Layer Sizes\end{tabular}} & \multicolumn{1}{c|}{} & \multicolumn{1}{c|}{\textbf{Parameter}}         & \multicolumn{1}{c|}{\textbf{Train $R^2$}} & \multicolumn{1}{c|}{\textbf{Test $R^2$}} & \multicolumn{1}{l|}{\textbf{\begin{tabular}[c]{@{}l@{}}Best Hidden\\ Layer Sizes\end{tabular}}} \\ \cline{1-4} \cline{6-9} 
\textbf{Reactivity}        & 0.9978               & 0.9914              & {[}256,128{]}                                                              & \multicolumn{1}{c|}{} & \multicolumn{1}{c|}{\textbf{R3 Last Pass BU}}   & \multicolumn{1}{c|}{0.9242}               & \multicolumn{1}{c|}{0.9019}              & \multicolumn{1}{l|}{{[}64{]}}                                                                   \\ \cline{1-4} \cline{6-9} 
\textbf{Power PC 1}        & 0.9931               & 0.9840              & {[}64{]}                                                                   & \multicolumn{1}{c|}{} & \multicolumn{1}{c|}{\textbf{R4 Last Pass BU}}   & \multicolumn{1}{c|}{0.9166}               & \multicolumn{1}{c|}{0.8975}              & \multicolumn{1}{l|}{{[}256{]}}                                                                  \\ \cline{1-4} \cline{6-9} 
\textbf{Power PC 2}        & 0.9955               & 0.9759              & {[}256,128{]}                                                              & \multicolumn{1}{c|}{} & \multicolumn{1}{c|}{\textbf{\# Discarded}}      & \multicolumn{1}{c|}{0.8463}               & \multicolumn{1}{c|}{0.7265}              & \multicolumn{1}{l|}{{[}32{]}}                                                                   \\ \cline{1-4} \cline{6-9} 
\textbf{Power PC 3}        & 0.9260               & 0.8410              & {[}32{]}                                                                   & \multicolumn{1}{c|}{} & \multicolumn{1}{c|}{\textbf{BU Bin 1}}          & \multicolumn{1}{c|}{0.9813}               & \multicolumn{1}{c|}{0.9807}              & \multicolumn{1}{l|}{{[}32{]}}                                                                   \\ \cline{1-4} \cline{6-9} 
\textbf{Power PC 4}        & 0.9909               & 0.9097              & {[}256,128{]}                                                              & \multicolumn{1}{c|}{} & \multicolumn{1}{c|}{\textbf{BU Bin 2}}          & \multicolumn{1}{c|}{0.9831}               & \multicolumn{1}{c|}{0.8430}              & \multicolumn{1}{l|}{{[}128,64{]}}                                                               \\ \cline{1-4} \cline{6-9} 
\textbf{Power PC 5}        & 0.9605               & 0.6608              & {[}256,128{]}                                                              & \multicolumn{1}{c|}{} & \multicolumn{1}{c|}{\textbf{BU Bin 3}}          & \multicolumn{1}{c|}{0.9348}               & \multicolumn{1}{c|}{0.7990}              & \multicolumn{1}{l|}{{[}128,8{]}}                                                                \\ \cline{1-4} \cline{6-9} 
\textbf{Flux PC 1}         & 0.9877               & 0.9726              & {[}256{]}                                                                  & \multicolumn{1}{c|}{} & \multicolumn{1}{c|}{\textbf{BU Bin 4}}          & \multicolumn{1}{c|}{0.9569}               & \multicolumn{1}{c|}{0.8210}              & \multicolumn{1}{l|}{{[}64{]}}                                                                   \\ \hline
\textbf{Flux PC 2}         & 0.9805               & 0.9490              & {[}256{]}                                                                  & \multicolumn{1}{c|}{} & \multicolumn{1}{c|}{\textbf{BU Bin 5}}          & \multicolumn{1}{c|}{0.9755}               & \multicolumn{1}{c|}{0.8295}              & \multicolumn{1}{l|}{{[}256,128{]}}                                                              \\ \hline
\textbf{Flux PC 3}         & 0.9871               & 0.9742              & {[}64,4{]}                                                                 & \multicolumn{1}{c|}{} & \multicolumn{1}{c|}{\textbf{BU Bin 6}}          & \multicolumn{1}{c|}{0.9460}               & \multicolumn{1}{c|}{0.7627}              & \multicolumn{1}{l|}{{[}128,64{]}}                                                               \\ \hline
\textbf{Flux PC 4}         & 0.9904               & 0.8628              & {[}256,128{]}                                                              & \multicolumn{1}{c|}{} & \multicolumn{1}{c|}{\textbf{BU Bin 7}}          & \multicolumn{1}{c|}{0.9152}               & \multicolumn{1}{c|}{0.8236}              & \multicolumn{1}{l|}{{[}64,4{]}}                                                                 \\ \hline
\textbf{Flux PC 5}         & 0.9698               & 0.8573              & {[}128,64{]}                                                               & \multicolumn{1}{c|}{} & \multicolumn{1}{c|}{\textbf{BU Bin 8}}          & \multicolumn{1}{c|}{0.9776}               & \multicolumn{1}{c|}{0.7891}              & \multicolumn{1}{l|}{{[}256,128{]}}                                                              \\ \cline{1-4} \cline{6-9} 
\textbf{Avg. Discharge BU} & 0.9859               & 0.9761              & {[}256{]}                                                                  & \multicolumn{1}{c|}{} & \multicolumn{1}{c|}{\textbf{BU Bin 9}}          & \multicolumn{1}{c|}{0.9661}               & \multicolumn{1}{c|}{0.9440}              & \multicolumn{1}{l|}{{[}16,8{]}}                                                                 \\ \cline{1-4} \cline{6-9} 
\textbf{R1 Last Pass BU}   & 0.9305               & 0.8910              & {[}64{]}                                                                   & \multicolumn{1}{c|}{} & \multicolumn{1}{c|}{\textbf{Avg. Power/pebble}} & \multicolumn{1}{c|}{0.9885}               & \multicolumn{1}{c|}{0.9896}              & \multicolumn{1}{l|}{{[}32,4{]}}                                                                 \\ \cline{1-4} \cline{6-9} 
\textbf{R2 Last Pass BU}   & 0.9332               & 0.8942              & {[}256{]}                                                                  &                       & \textbf{}                                       &                                           &                                          &                                                                                                 \\ \cline{1-4}
\end{tabular}
\centering
\label{tab:Rsquared_values_and_hyperparams}
\end{table*}

\subsection{Model Performance}

To assess the performance of the LSTM, the coefficient of determination, $R^2$, is calculated for each variable using the training and test datasets in Table~\ref{tab:Rsquared_values_and_hyperparams}. A value of 1 means that the model can explain all the variation in the target variable, where 0 or less suggests that using the average is more accurate. 

\begin{figure}[htb]
\centering
\begin{subfigure}[b]{0.9\columnwidth}
\centering
\includegraphics[width=\textwidth]{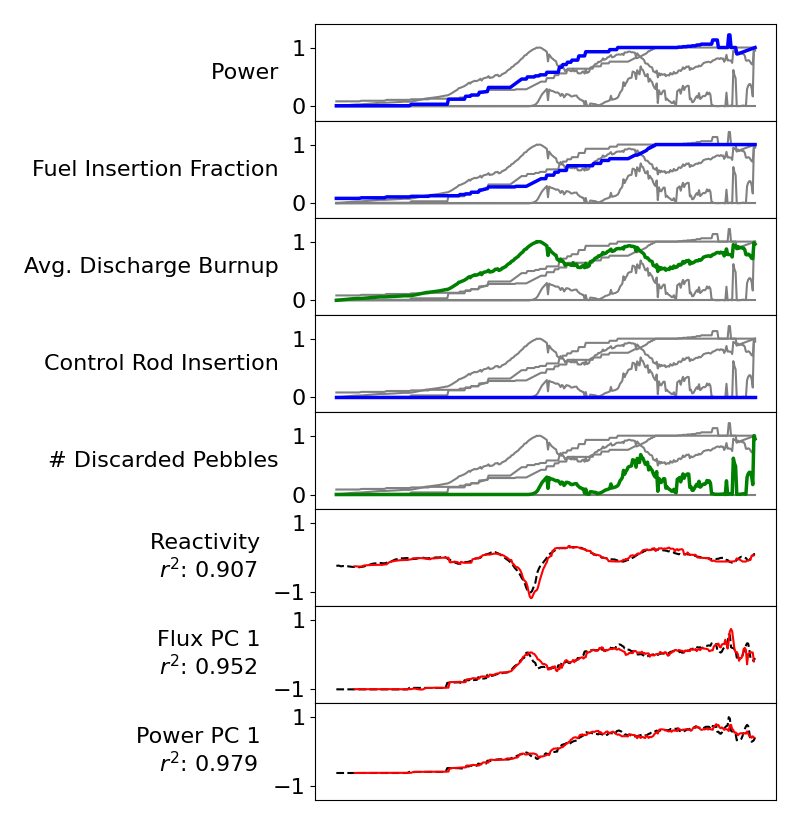}
\end{subfigure}

\begin{subfigure}[b]{0.9\columnwidth}
\centering
\includegraphics[width=\textwidth]{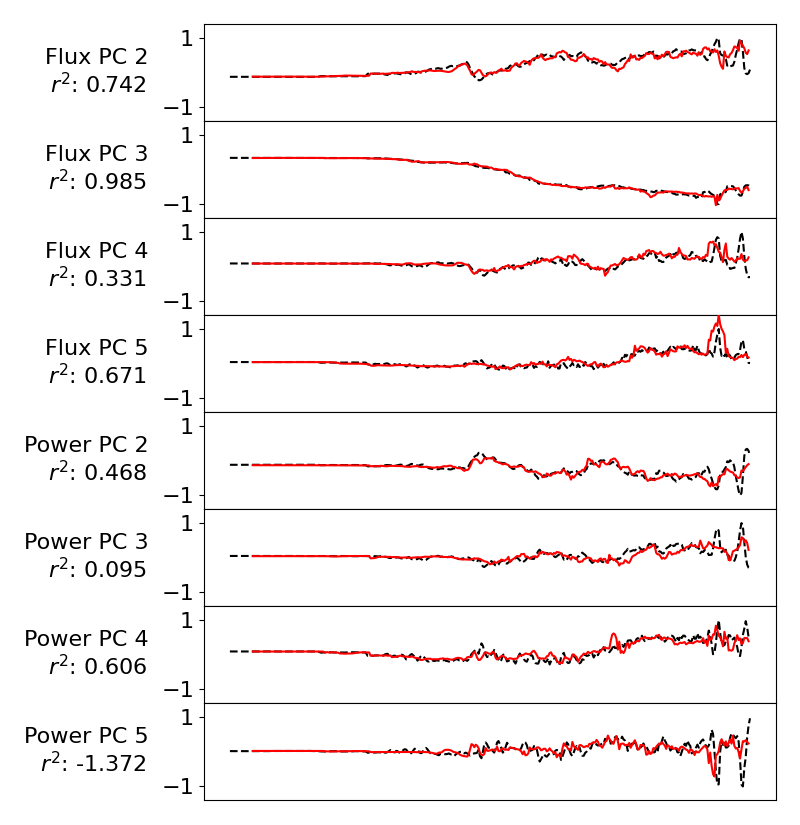}
\end{subfigure}
\caption{Performance of the LSTM on an entirely unseen sequence.}
\label{fig:sequence_prediction_performance}
\end{figure}

The performance for reactivity and lower rank flux and power components is satisfactory, with the model being able to successfully track oscillations in these variables due to fuel discharge. Predicting higher rank mesh components becomes more troublesome, as these variables represent increasing amounts of statistical noise in the mesh. This makes them difficult to relate to physical quantities. This motivates potentially excluding them if they worsen the mesh reconstruction.

The prediction capabilities of the LSTM for the held out sequence is shown in Figure~\ref{fig:sequence_prediction_performance}. This sequence, while simple, was completely excluded from the training and hyperparameter tuning process. General trends in reactivity and the lower principal components are well captured, although finer perturbations are not captured as well.


\begin{figure*}[p]
\centering
\begin{subfigure}[b]{0.34\textwidth}
    \centering
    \includegraphics[width=\textwidth]{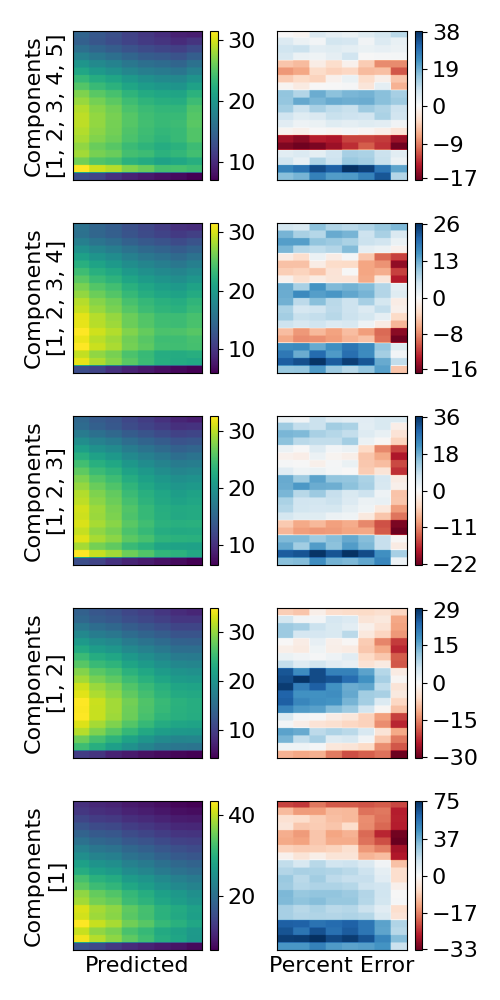}
    \label{fig:mesh_reconstruction:parta}
\end{subfigure}
\begin{subfigure}[b]{0.34\textwidth}
    \centering
    \includegraphics[width=\textwidth]{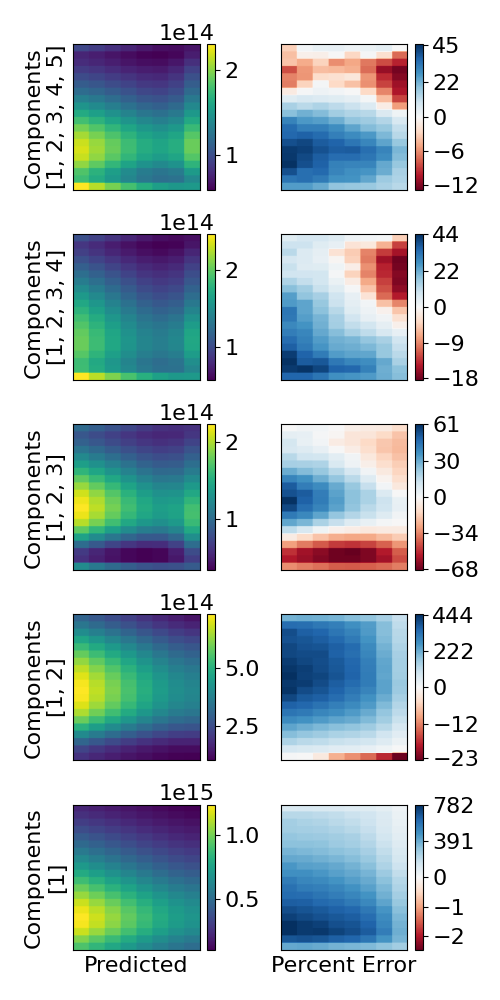}
    \label{fig:mesh_reconstruction:partb}
\end{subfigure}
\caption{Reconstruction of the power mesh (left) and the thermal flux mesh (right) a single timestep at equilibrium. The cell-wise percent error difference between the reconstructed mesh and ground truth mesh is shown. This same meshes as shown in Figure~\ref{fig:mesh_examples} are compared to.}
\label{fig:mesh_reconstruction}
\end{figure*}

\begin{figure*}[p]
\centering
  \begin{subfigure}[b]{0.35\textwidth}
    \centering
    \includegraphics[width=\textwidth]{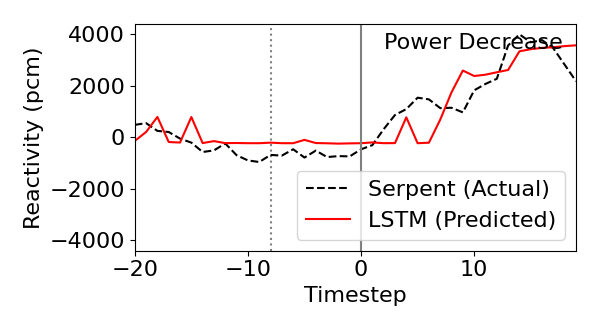}
    \label{fig:forecastperformance:power_down}
  \end{subfigure}
  \begin{subfigure}[b]{0.35\textwidth}
    \centering
    \includegraphics[width=\textwidth]{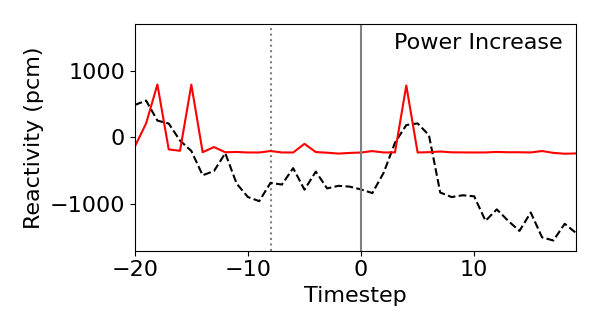}
    \label{fig:forecastperformance:power_up}
  \end{subfigure}
  
  \begin{subfigure}[b]{0.35\textwidth}
    \centering
    \includegraphics[width=\textwidth]{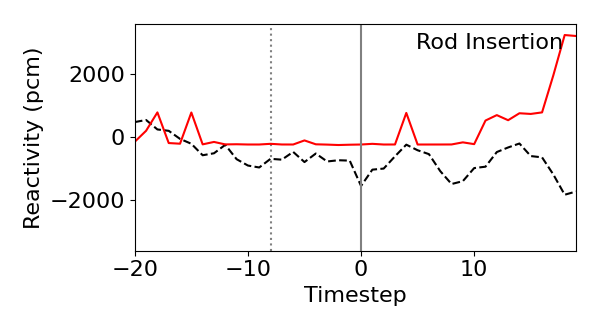}
    \label{fig:forecastperformance:rods_in}
  \end{subfigure}
  \begin{subfigure}[b]{0.35\textwidth}
    \centering
    \includegraphics[width=\textwidth]{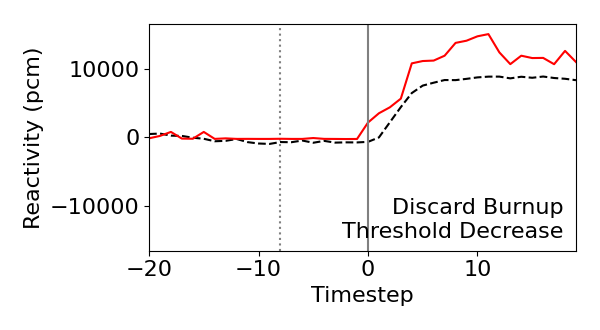}
    \label{fig:forecastperformance:threshold_decrease}
  \end{subfigure}
  \caption{Comparsion of reactivity between PEARLSim ground truth and LSTM forecast. The 8 timestep window is shown.}
  \label{fig:forecastperformance}
\end{figure*}

\subsection{Mesh Reconstruction}
The power and flux meshes are reconstructed from their predicted principal components. The percent error of each cell of the mesh from the original Serpent-generated mesh is calculated. This is shown for the power and thermal flux meshes in Figures~\ref{fig:mesh_reconstruction}. The results are compared using different combinations of the principal components. When a component is left out, its values are zeroed before the inverse transformation is applied. This removes its influence from the reconstructed mesh.

For power, it appears the mesh reconstruction is slightly less accurate when the fifth principal component is included in the calculation, since it likely only captures noise from the statistical uncertainty of the Serpent tally. For flux, leaving out components can introduce massive amounts of error. This is likely because PCA is applied to flux in all three energy groups, which are strongly related but evolve differently as the core is operated.

\subsection{Forecasting}

The ability to forecast changes to reactivity that would result from an operation sequence is demonstrated. A simple run-in is simulated, with both the power and the fuel pebble insertion ratio quickly reaching 100\% in large steps. This type of sequence is well captured in the training and is the same for all forecasting scenarios. This part of the sequence is shown in Figure~\ref{fig:forecastperformance} as occurring before timestep 0.

Each of the input controls are then independently perturbed by a substantial amount, and then held constant for 20 time steps. This is simulated with PEARLSim to get the ground truth. Meanwhile, it is predicted with the LSTM using only the control and dependent input history up to time step 0. The comparison of the two models is shown in Figure~\ref{fig:forecastperformance}.

Generally, the model performs worse the further into the future it predicts. This is because the error accumulates with repeated calculations of dependent variables. This process can help identify which operating parameters are well represented in the training data and thus well predicted by the model. Decreasing the power and burnup discard threshold at equilibrium are both well predicted and well represented in the data set. However, most of the run-in sequences largely emphasize control rod removal rather than insertion. This type of assessment can serve as another metric for how well the model generalizes.

\FloatBarrier
\section{Run-in Optimization}
\label{sec:startupautomation}

\subsection{Algorithm Overview}

The trained LSTM can be used as a simple artificial intelligence controller to guide a core simulator through the running-in phase of operation. This is done by combining a set of operation constraints with the ability to predict the impact on reactivity from perturbing different controls. A short sequence of goal control states can be specified for the simulator to be guided towards. The initial and final goal states are shown in Table~\ref{tab:run-in-parameters}. However, intermediate goals could be added if motivated by testing or fuel qualification needs. 

\begin{table}[!htb]
\centering
\caption{Goal points and perturbation values used in running-in calculations. Each control has a grid of 1,000 possible values on the way to their final values.}
\begin{tabular}{|l|l|l|l|}
\hline
\textbf{Parameter}                                                                     & \textbf{\begin{tabular}[c]{@{}l@{}}Starting \\ Value\end{tabular}} & \textbf{Perturbation} & \textbf{\begin{tabular}[c]{@{}l@{}}Final \\ Value\end{tabular}} \\ \hline
\textbf{Power (kW)}                                                                    & 10                                                                 & +279.99               & 280,000                                                         \\ \hline
\textbf{\begin{tabular}[c]{@{}l@{}}Graphite pebbles\\ insertion fraction\end{tabular}} & 0.8879                                                             & -$8.879\times10^{-4}$ & 0                                                               \\ \hline
\textbf{\begin{tabular}[c]{@{}l@{}}Control rod\\ depth (cm)\end{tabular}}           & 60.25                                                              & +0.30922              & 369.47                                                          \\ \hline
\textbf{Burnup step (d)}                                                               & 6.525                                                              & $\pm$0.01305          & 6.525                                                           \\ \hline
\end{tabular}
\label{tab:run-in-parameters}
\end{table}

The model perturbs each operating parameter towards its goal value, and the LSTM predicts whether this will have a positive or negative effect on reactivity. It then fine-tunes the parameters to achieve reactivity as close to 0 possible, with a tolerance set by the user; 50 pcm is used for this study. PEARLSim is subsequently run with the chosen set of parameters, simulating the next step towards the goal state and generating updated values for the dependent input variables. These values are then fed back to the model. If the model's prediction is perfectly correct, then resulting Serpent calculation will also have near-zero reactivity. A minimum number of control perturbations, $s$, is imposed to govern the overall rate at which the model approaches its target state.

\subsection{Performance}

\begin{figure}[htb]
\centering
\begin{subfigure}[b]{0.4\textwidth}
    \centering
    \includegraphics[width=\textwidth]{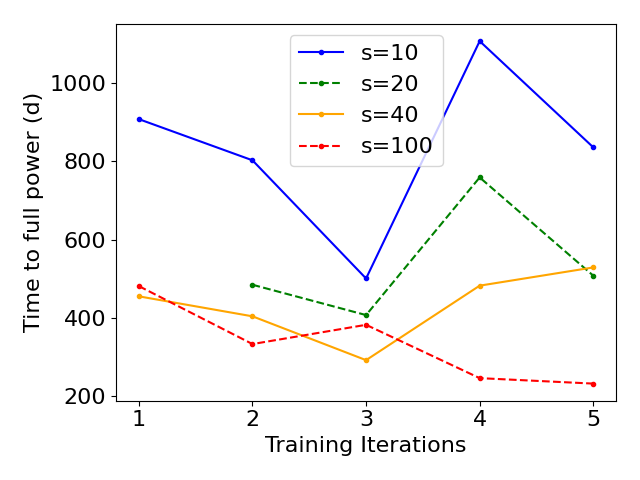}
    \label{fig:run_in_optimization:parta}
\end{subfigure}

\begin{subfigure}[b]{0.4\textwidth}
    \centering
    \includegraphics[width=\textwidth]{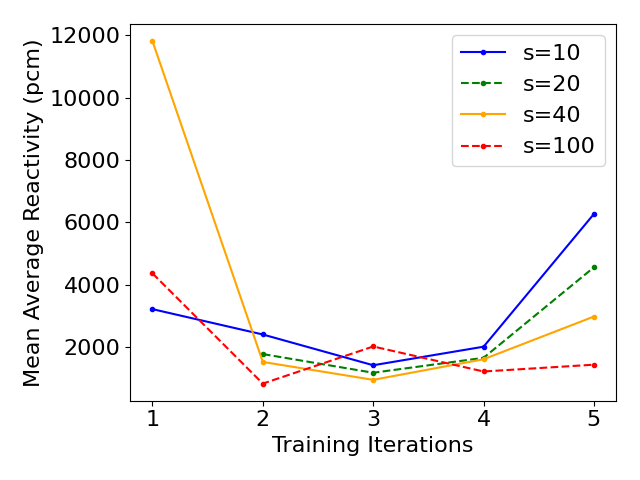}
    \label{fig:run_in_optimization:partb}
\end{subfigure}
\caption{Performance of the LSTM for running-in optimization shown over multiple training cycles. The time it takes to get to full power is shown (top), as well as the mean absolute error in reactivity (bottom).}
    \label{fig:run_in_optimization}
\end{figure}

This simulation process is iterative. For every attempted run-in simulation, PEARLSim generates more data which can be used to retrain and improve the model. The gradual improvement of the model is demonstrated in Figure~\ref{fig:run_in_optimization} for different values of s. Because the training process for the LSTM is stochastic, the trained model can sometimes perform worse.

For each iteration, it was predicted whether changing power, withdrawing control rods, and inserting less graphite pebbles would produce a positive or negative effect on reactivity. If additional negative reactivity was needed at any time, then the circulation rate is decreased to allow the pebble burnup to increase. 

\begin{figure}[!htb]
  \centering
  \includegraphics[width=0.95\columnwidth]{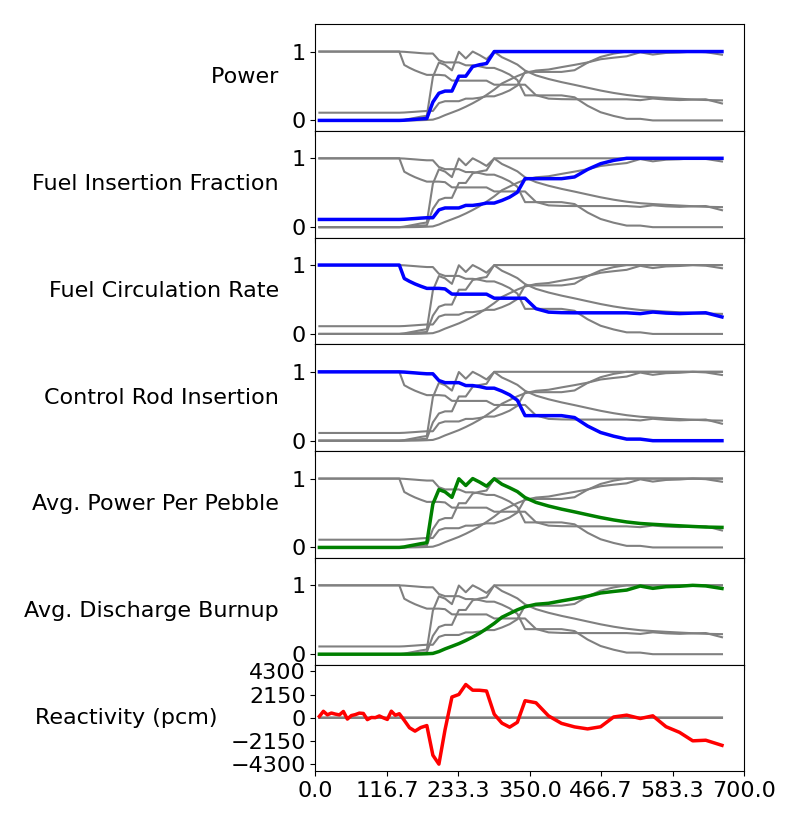}
  \caption{Example of a running-in simulation guided by the LSTM, taken from the 3rd training iteration with $s=$40.}
  \label{fig:run-in-example}
\end{figure}

An example of a run-in sequence that performed comparatively well is shown in Figure~\ref{fig:run-in-example}. It can be seen that gradually increasing the power and fraction of fuel pebbles inserted in tandem keeps the power per pebble from becoming too large. This also helps achieve a more even burnup distribution. If the core reaches full power with a burnup distribution that is skewed towards the discard threshold, it will experience a massive reactivity increase when those pebbles are removed and replaced. 

This workflow establishes a reinforcement-based feedback loop for the LSTM, in which the model effectively determines where to collect new data and improve its understanding of the underlying reactor physics.

\section{Conclusion}
\label{sec:conclusion}

LSTMs are a powerful tool for predicting the complex, time dependent behavior of PBRs. The models have demonstrated very good performance for determining reactivity and PCA-reduced power and flux meshes over large operation sequences that are either simple, realistic, or highly randomized. They are capable of forecasting future values of target variables as well as dependent input variables but could benefit from additional data to improve its generalization. The models have also been shown to offer a powerful tool for steering core simulations through the running-in phase, which could help PBR developers minimize the amount of time that is spent at low power.

Some changes to PEARLSim could improve the quality of the data generated and thus allow for more precise and smooth predictions of the target variables. For example, shuffling of pebble locations likely produces a significant amount of variance in reactivity. This can be combated by running multiple transport simulations per step with the pebbles shuffled differently each time and averaging the target variables. Due to the significant increase in computing time, this would require very purposeful planning of the training sequences. More particles or cycles can be used in Serpent to reduce the statistical uncertainty on the power and flush mesh, which will especially help for predicting higher-order principal components. Coupling PEARLSim and Serpent with a thermal hydraulics code is also very important, as it would make the reactivity feedback from power increases more realistic as the temperature in the core changes. 

Beyond this, there are ways the machine learning infrastructure discussed here can be improved. Additional feature engineering can be done to improve model generalization. Physics-based features that can be calculated quickly or deterministically and used as a dependent input could be helpful. This could include a $k_{\infty}$ feature calculated with a reduced order model, or other quantities that can be derived from bulk pebble-wise measurements. The use of other model types, such as Gated Recurrent Units (GRU) or autoencoders, should be explored. 

Finally, the use of multiple types of fuel with different levels of enrichment should be explored with this methodology. Faster run-in sequences are likely possible if lower enrichment fuel pebbles are also used, although their inclusion would make the process even more complicated.

\section*{Acknowledgments}

This research uses the Savio computational cluster resource provided by the Berkeley Research Computing program at the University of California, Berkeley (supported by the UC Berkeley Chancellor, Vice Chancellor for Research and Chief Information Officer).

\bibliography{bibliography}

\end{document}